\documentclass[twocolumn,aps,showpacs]{revtex4-1}
\usepackage{makeidx}
\usepackage{bm}
\usepackage{amssymb}
\usepackage{amssymb,amsmath}
\usepackage{graphicx}
\usepackage{graphics}
\usepackage{epstopdf}
\usepackage{graphicx}
\usepackage{graphics}
\usepackage{dsfont}
\usepackage{slashed}
\usepackage{ulem}

\newcommand{\PathToFigures}{}
\newcommand{\nn}{\nonumber}

\begin{document}
\title{ Critical behavior of the 2D Ising model with
long-range correlated disorder  }
\author{M. Dudka$^1$, A. A. Fedorenko$^2$, V. Blavatska$^1$ and  Yu. Holovatch$^1$}
\affiliation{ $^1$Institute for Condensed Matter Physics of the National
Academy of Sciences of Ukraine, 79011 Lviv, Ukraine \\
$^2$\mbox{Univ Lyon, Ens de Lyon, Univ Claude Bernard, CNRS, Laboratoire de Physique, F-69342 Lyon, France}}

\date{June 22, 2016}


\begin{abstract}
We study  critical behavior of the diluted 2D
Ising model in the presence of disorder
correlations which decay algebraically with distance as $\sim r^{-a}$. Mapping the problem
onto  2D Dirac fermions with correlated disorder we calculate the critical properties using renormalization group
 up to two-loop order. We show that beside the Gaussian fixed point the flow equations have a non trivial fixed point which is
stable for $0.995<a<2$ and is characterized by the correlation
length exponent $\nu= 2/a + O((2-a)^3)$. Using bosonization, we also
calculate the averaged square of the spin-spin correlation function
and find the corresponding critical exponent $\eta_2=1/2-(2-a)/4+O((2-a)^2)$.

\end{abstract}

\maketitle

\section{Introduction}

Effects of quenched disorder on critical behavior attracted
considerable attention for several
decades~\cite{stichcombe-83,chardy-96,pelissetto-02,alonso-01,holovatch-02,grinstein-76}.
Among various aspects of this problem influence of disorder
correlations is of particular interest.  Examples include spin
models with correlated random
bonds~\cite{McCoy68,weinrib-83,dorogovtsev-80,decesare-94,korzhenevskii-94,korucheva-98,fedorenko-00,Blavatska2005}
and random
fields~\cite{Vojta1995,fedorenko2007,Ahrens2011,Baczyk2013}, quantum
transport and localization
\cite{Croy2011,fedorenko2012,andreanov2014}, polymers in random
media~\cite{Blavatska2001}, disordered elastic systems
\cite{Fedorenko2006,Fedorenko2008}, and
percolation~\cite{Schrenk2013}.

The 2D Ising model is historically important  for studying criticality since its critical behavior
deviates from the mean-field picture but still allows for an exact solution~\cite{onsager}.
According to the  Harris criterion~\cite{harris-74} uncorrelated random bond or random site
disorder modifies the critical behavior provided that
the heat capacity exponent of the pure system is positive, $\alpha_{\rm pure}>0$.
Although uncorrelated  disorder is only
marginally irrelevant for the 2D Ising system, since $\alpha_{\rm pure}=0$,  its effects on the critical behavior were a subject of intensive theoretical and numerical studies~\cite{2DRIM}.
Apart from  the purely academic interest, this problem has  potential applications; \textit{e.g.}, it was observed that domain formation in
membranes with quenched protein obstacles without preferred affinity can be described by a diluted 2D Ising model~\cite{Fischer11}.

The solution of the pure 2D Ising model can be formulated in terms of free 2D Majorana fermions
whose mass is proportional to the reduced temperature~\cite{schultz64}.
The presence of disorder adds a four-fermion interaction with the coupling constant proportional to the concentration
of impurities~\cite{dotsenko83,dotsenko81,dotsenko82}. The resulting model has been intensely studied
by renormalization group methods. These studies not only confirmed the
marginal irrelevance of the disorder but also revealed the presence of
logarithmic corrections to the critical behavior
of the pure model.
In particular, it was found that the
specific heat singularity modifies from $C\sim\ln(1/\tau)$ to  $C\sim\ln\ln(1/\tau)$
where $\tau = (T_c- T)/T_c$ if the temperature  goes sufficiently close to the
critical temperature $T_c$~\cite{dotsenko83,dotsenko81}. The calculation of the correlation function is a much more difficult
task since in the fermionic picture the spin operator is a nonlocal object so that even
for the pure case it requires some efforts to recover the well-known result
$\eta_{\mathrm{pure}}=\frac14$.
Initially it was argued~\cite{dotsenko83,dotsenko82} that disorder modifies the critical exponent to $\eta=0$, but later
it was realized that the behavior of the  $N$th moment of the spin-spin correlation
function averaged over disorder configurations is~\cite{shankar87,ludwig90}
\begin{equation}
\overline{ G(r)^N}
\sim \frac{(\ln r)^{N(N-1)/8}}{r^{N/4}},
\end{equation}
while in the pure model $ G(r)^N \sim  r^{-N/4}$.

Real systems may contain extended defects such as linear dislocations or
grain boundaries which are
either aligned in space or may have random orientation.
The presence of extended defects or long-range (LR) correlated disorder
modifies the Harris criterion opening a possibility for relevance of disorder
in two dimensions. Almost a half century ago, McCoy and Wu  proposed  the disordered 2D Ising model in which impurities are perfectly correlated in one direction and
uncorrelated in the transverse direction~\cite{McCoy68}. Though it was originally argued that the phase
transition in this model is smeared, later it was
shown that it is sharp but controlled by an infinite-randomness
fixed point \cite{fisher92}.
An extension of this model to $d$ dimensions  was proposed in
Ref.~\cite{dorogovtsev-80}, where extended defects are  infinitely
correlated in $\varepsilon_d$ dimensions and randomly distributed in
the remaining $\tilde{d} =d-\varepsilon_d$ dimensions. Values
$\varepsilon_d=0$, $1$, $2$ correspond to uncorrelated point-like,
linear and planar defects, respectively, while non-integer values of
$\varepsilon_d$  may describe systems containing fractal-like
defects~\cite{yamazaki-88}. The critical equilibrium and dynamic
behavior of these and related models were studied using a double
expansion in $\varepsilon = 4-d$ and $\varepsilon_d$
in Refs. \cite{boyanovsky-82,prudnikov-83,lawrie-84,blavatska-03,%
decesare-94,korzhenevskii-94,yamazaki-86,yamazaki-88,fedorenko04}. The numerical
studies of systems
with parallel linear~\cite{Vasiliev15} and planar defects~\cite{lee-92,vojta-03}
were also performed.

Weinrib and Halperin proposed an alternative model~\cite{weinrib-83}
with LR correlated  disorder whose correlations decay with the
distance $r$ as a power-law, $g(r) \propto r^{-a}$. The critical
behavior of this model has been studied  to two-loop order using a
double $\varepsilon=4-d$, $\delta=4-a$ expansion~\cite{korucheva-98}
and also direct calculations in $d=3$~\cite{fedorenko-00}. These
studies suggest that the phase transition belongs to a universality
class different from that for systems with uncorrelated disorder  if
the correlation length  exponent of the pure (undiluted) model
satisfies $\nu_{\rm pure} < 2/a$. The condition holds for $a<d$,
while for $a>d$ the usual Harris criterion \cite{harris-74} is
recovered and this condition is substituted by   $\nu_{\rm pure} <
2/d$. Although results of Refs.~\cite{weinrib-83,korucheva-98,fedorenko-00} are in qualitative
agreement and predict an emergence of the new type of critical
behavior governed by the so-called LR disorder fixed point, they do
not agree on quantitative level. In particular, results of
Refs.~\cite{weinrib-83,korucheva-98} suggest that in the new
universality class the correlation length  exponent is $\nu=2/a$ to
the second order in $\varepsilon=4-d$ and $\delta=4-a$ (and even probably to all orders,
see \cite{weinrib-83,HonkonenNalimov}), whereas calculations performed
directly in
three dimensions~\cite{fedorenko-00} are in favor of a non-trivial
value of the exponent, which differs from $\nu=2/a$ already in the
two-loop approximation.
In principle the discrepancy  can be explained by breaking down
the $\varepsilon=4-d$-expansion at large $\varepsilon$.
In order to verify this conjecture one needs a controllable method which does not rely  on $\varepsilon=4-d$ expansion
with analytical continuation to $\varepsilon=2$.
Subsequently, these analytic results have
been checked by numerical
calculations~\cite{ballesteros99,prudnikov-05,ivaneyko-08,Bagamery2005}. In turn, these have not
 led so far to common agreement either.
Results of computer simulations in Ref.~\cite{ballesteros99,Bagamery2005} support
the analytic result
$\nu=2/a$, whereas the critical exponents obtained in numerical studies
in Refs.~\cite{prudnikov-05,ivaneyko-08} deviate from this prediction raising the
question about
dependence of the critical exponents on the peculiarities of disorder distribution.

In this paper we reconsider this problem using mapping of the 2D Ising model with
LR correlated disorder to  disordered 2D Dirac fermions, and thus, approaching the
problem from low dimensions. This has been done to one-loop order in
Refs.~\cite{rajabpour08,fedorenko2012}.
We extend these calculations to two-loop order and also compute
the averaged square of the spin-spin correlation function to the lowest order
using bosonization.
Since the calculations are done directly in two dimensions and are well controlled in the limit of small $\delta=2-a$
they provide a test for the possible breaking down
of the $\varepsilon=4-d$ expansion.

The rest of the paper is
organized as follows: Section~\ref{sec:model} introduces
fermionic representation of the 2D Ising model with correlated disorder. We give a short description of renormalization of this model
 in Sec.~\ref{III}. We present two-loop scaling functions in Sec.~\ref{IV} together
 with their analysis
 within the framework of $\delta$-expansion.
Section~\ref{V} is devoted to calculation of the averaged square of the spin-spin correlation function using mapping
to the sine-Gordon model.  We end the paper with  conclusions in Sec.~\ref{VI}. Some technical points are given in the Appendices.

\section{Model} \label{sec:model}

The random bond 2D Ising model can be described by two-dimensional real Majorana
fermions whose action reads~\cite{Itzykson-Drouffe-book}
\begin{eqnarray}
S_M= \int d \bar{z} d z\, \left[ {\chi} \bar{\partial} \chi  +
\bar{\chi} \partial \bar{\chi} +
i m(z) \bar{\chi} \chi \right], \label{eq:action1}
\end{eqnarray}
where $\bar{\chi}(z)$ and ${\chi}(z)$ are one-component Grassmann fields,
$z=x+iy$, $\partial = \frac12(\partial_x-i\partial_y)$,
and $m(z)= m_0 + \delta m(z) $ is coupled to the energy operator
$\epsilon(z)=i \bar{\chi}(z) {\chi}(z) $.
Here $m_0 = (T_c- T)/T_c$ and $\delta m(z)$ encodes
spatial variations in bond strength for a given realization of disorder.
Using the two-component spinor notation $\Psi = (\chi,\bar{\chi})^{T}$
action~(\ref{eq:action1}) can be rewritten as
\begin{eqnarray}
S_M=\frac{1}{2} \int d^2 r \, \bar{\Psi}(r)\left[\slashed{\partial} +
  m(r) \right]\Psi(r),
\end{eqnarray}
where $\slashed{\partial} = \gamma_j \partial^j$ with $\gamma_j=\sigma_j$
($j=1,2$) being the Pauli matrices .
Note that $\bar{\Psi}$ is not an independent field, it is related to $\Psi$
by $\bar{\Psi}=\Psi^T \gamma_0 $ with $\gamma_0=\sigma_2$.
We assume that $\delta m(r)$  is a Gaussian  random variable with zero mean and
a variance decaying as a power law
\begin{eqnarray} \label{eq:dis-cor-0}
\overline{\delta m(r)\delta m(0) }= g(r) \sim r^{-a},\,\,\,\,\,r\to\infty.
\end{eqnarray}
To simplify calculations we follow~\cite{shankar87} and introduce two
Majorana  fermions  $\Psi_1$ and $\Psi_2$  which combine to form a
complex Dirac fermion $\psi = (\Psi_1 + i \Psi_2)/\sqrt{2} $.
The corresponding action reads
\begin{eqnarray} \label{eq:action-Dirac}
S_D=  \int d^2 r \, \bar{\psi}(r)\left[\slashed{\partial} +
  m(r) \right]\psi(r).
\end{eqnarray}
Note  that $\bar{\psi}$ and $\psi$ are independent and we may
change variable $\bar{\psi} \to - i \bar{\psi}$.  Then the resulting action
at criticality, $m_0=0$, corresponds to the Dirac fermions in the presence of random imaginary
chemical potential $-i \delta m(r)$.  Changing variable  $\bar{\psi} \to  - \bar{\psi} \sigma_3 $
one can see that action~(\ref{eq:action-Dirac}) also describes the 2D Dirac fermions
with random mass disorder~\cite{ludwig94}.

In what follows we are going to use dimensional regularization. To
that end we have to generalize the problem to arbitrary $d$ and replace
the Pauli matrices by a Clifford algebra represented by the matrices $\gamma_i$
satisfying the anticommutation relations~\cite{zinn-justin-book}:
\begin{eqnarray} \label{eq:com-rel}
\gamma_i\gamma_j + \gamma_j \gamma_i = 2 \delta_{i j} \mathbb{I}, \ \ \ \ \
i,j=1,...,d.
\end{eqnarray}
To average over disorder
we use the replica trick
introducing $n$ copies of the original system~\cite{edwards_replica}.
The resulting replicated action reads
\begin{eqnarray} \label{eq:action}
&& \!\!\!\!\!\!\!\! S = - i \sum\limits_{\alpha=1}^n
\int d^d r  \bar{\psi}_{\alpha}({r})(\slashed{\partial}  + m_0  )
  \psi_{\alpha}({r}) \nonumber \\
&& \!\!\!\!\!\!\!\!
 + \frac12 \sum\limits_{\alpha,\beta=1}^n
\int d^d {r} d^d {r}' g({r}-{r}')
\bar{\psi}_{\alpha}({r}) \psi_{\alpha}({r})
\bar{\psi}_{\beta}({r}') \psi_{\beta}({r}').\ \ \
\end{eqnarray}
The properties of the
original system with quenched disorder are then obtained by taking
the limit $n\to 0$.
It is convenient to fix the normalization of the disorder distribution~(\ref{eq:dis-cor-0}) in Fourier space.
We  take disorder potential to be random Gaussian with zero mean and the correlator
\begin{eqnarray}
\overline{\delta m(k)\delta m(k') }= (2\pi)^d \delta^d(k+k') g(k).
\end{eqnarray}
We choose
\begin{eqnarray}\label{furier}
g(k) = u_0 + v_0 k^{a-d},
\end{eqnarray}
here $u_0$ and $v_0$ are bare coupling constants.
The LR coupling constant $v_0$ is  relevant only for $a<d$.  Note that if one neglects the SR term $u_0$
in Eq.~(\ref{furier}) it will be ultimately generated by the RG flow.
The bare propagator of the action~(\ref{eq:action}) can be written as
\begin{eqnarray}
\langle \bar{\psi}_\alpha({k}) \psi_{\beta}(-{k}) \rangle_0
=\delta_{\alpha\beta}
\frac{\gamma_j k_j  + i m_0}{k^2+m_0^2}. \ \label{eq:bare-propagator}
\end{eqnarray}

\section{\label{III} Renormalization of the model }

Using the bare propagator (\ref{eq:bare-propagator}) one can calculate the correlation
functions for the  action (\ref{eq:action})  perturbatively in $u_0$ and $v_0$.
The integrals entering  this perturbation series turn out to be ultraviolet (UV)
divergent in $d=2$.
To make the theory finite we are using the dimensional regularization~\cite{HooftVelman} and compute all integrals in $d=2-\varepsilon$.
Following the works \cite{weinrib-83,fedorenko2012} we perform a double expansion in $\varepsilon=2-d$  and $\delta=2-a$ so that all divergences
are transformed into the poles in $\varepsilon$  and $\delta$ while the ratio $\varepsilon/\delta$ remains finite.
In the framework of the minimal subtraction scheme we do not include these finite ratios
into the counterterms choosing them to be the pole part only. We are interested in the case  $0<a<2$, so that $0<\delta<2$, however,
one has to take with caution the numerical estimations for $\delta>1$ computed using the results obtained perturbatively in $\delta$.
We define the renormalized fields $\psi$, $\bar{\psi}$, mass $m$, and dimensionless coupling constants $u$ and $v$
in  such a way that all poles can be hidden in the renormalization factors $Z_\psi$, $Z_m$, $Z_u$ and $Z_v$ leaving finite
the correlation functions computed with the renormalized action
\begin{eqnarray} \label{eq:action-r}
&& \!\!\!\!\!\!\!\! S_R = \sum\limits_{\alpha=1}^n
\int_k \bar{\psi}_{\alpha}(-{k})(  Z_\psi \gamma_j k_j  - Z_m i m  )
  \psi_{\alpha}({k}) \nonumber \\
&& \!\!\!\!\!\!\!\!
+ \frac12 \sum\limits_{\alpha,\beta=1}^n \int_{k_1,k_2,k_3} \,
\left[ \mu^{\varepsilon} Z_u u + \mu^{\delta} Z_v v |{k}_1+{k}_2|^{a-d} \right] \nonumber \\
&& \!\!\!\!\!
\times \bar{\psi}_{\alpha}({k}_1) \psi_{\alpha}({k}_2)
\bar{\psi}_{\beta}({k}_3) \psi_{\beta}(-{k}_1-{k}_2-{k}_3),
\end{eqnarray}
where $\int_k := \int \frac{d^d k}{(2\pi)^d}$ and we have introduced a renormalization scale $\mu$.
Since the renormalized action is obtained from the bare one by the fields rescaling
\begin{eqnarray}
\psi_0 = Z_{\psi}^{1/2} \psi, \ \ \  \bar{\psi}_0 = Z_{\psi}^{1/2} \bar{\psi},
\end{eqnarray}
the bare and renormalized parameters are related by
\begin{eqnarray}
m_0 &=& {Z_m}Z_{\psi}^{-1} m,  \\
u_0 &=&  \mu^{\varepsilon }  Z_u Z_{\psi}^{-2} u,\ \ \ \
v_0 = \mu^{\delta }  Z_v Z_{\psi}^{-2} v,
\end{eqnarray}
where we have included $K_d/2$ in redefinition of $u$ and $v$.
$K_d = 2\pi^{d/2}/((2\pi)^d\Gamma(d/2))$ is the surface area of
the $d$-dimensional unite sphere divided by $(2\pi)^d$.
The renormalized $\mathcal N$-point vertex function
${\Gamma}^{(\mathcal N)}$  is related to the bare
$\mathring{\Gamma}^{(\mathcal N)}$ by
\begin{equation}
\mathring{\Gamma}^{(\mathcal N)}(k_i; m_0,u_0,v_0) = Z_{\psi}^{-{\mathcal N}/2}{\Gamma}^{(\mathcal N)}(k_i;  m ,u,v, \mu). \ \
\end{equation}
To calculate the renormalization constants it is enough to renormalize the two-point vertex function $\Gamma^{(2)}$ and
the four-point vertex function $\Gamma^{(4)}$. We impose that they are finite at $m=\mu$ and find the renormalization constants
using minimal subtraction scheme~\cite{Hooft}.  To that end it is convenient to
split the four-point function in the short-range (SR) and long-range (LR) parts:
\begin{eqnarray}
\Gamma^{(4)}(k_1,k_2,k_3,k_4){=}\Gamma_u^{(4)}(k_i)+\Gamma_v^{(4)}(k_i)|k_1+k_2|^{a-d}.
\end{eqnarray}
The renormalization constants are determined from the condition that
$\Gamma_u^{(4)}(0;m=\mu)$ and $\Gamma_v^{(4)}(0;m=\mu)$  are finite.

\begin{figure}
\includegraphics[width=80mm]{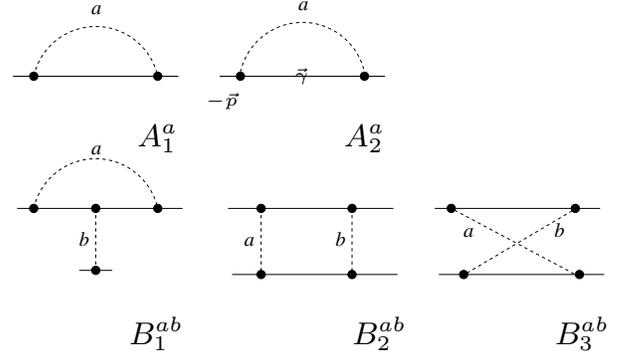}
\caption{The one-loop diagrams contributing to the two-point vertex function $\Gamma^{(2)}$  (first row) and to the four-point vertex function $\Gamma^{(4)}$ (second row) in the replica limit $n\to 0$. Solid lines correspond to the propagator~(\ref{eq:bare-propagator}) and dashed lines to the disorder
vertex~(\ref{furier}) which is split in the SR and LR parts.
The indices $a,\,b,\,c$  take values 0 or 1, depending on whether the dashed line stands for $u$-vertex or $v$-vertex.}
  \label{fig:one-loop-diagrams}
\end{figure}
\begin{figure}
\includegraphics[width=80mm]{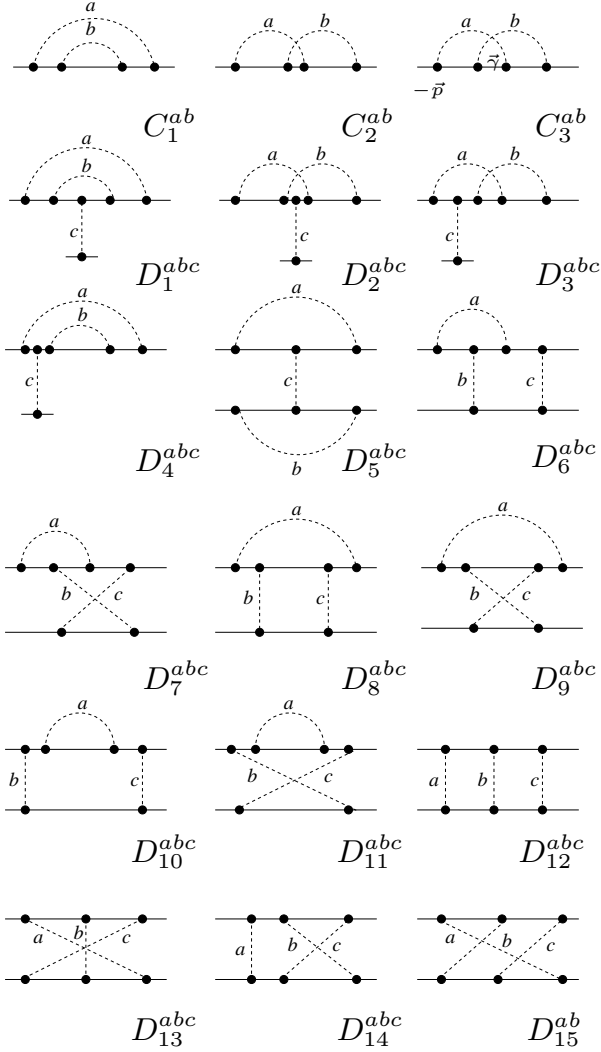}
\caption{The  two-loop diagrams contributing  to the two-point
vertex function $\Gamma^{(2)}$  (first row) and to the four-point
vertex function $\Gamma^{(4)}$  in the replica limit $n\to 0$. The indices $a,\,b,\,c$  take values
0 or 1, depending on whether the dashed line stands for $u$-vertex or
$v$-vertex.  }
  \label{fig:two-loop-diagrams}
\end{figure}

Since the bare vertex function does not depend on the renormalization scale $\mu$
the renormalized vertex function satisfies the renormalization group equation
\begin{eqnarray}
&&\left[\mu\frac{\partial}{\partial \mu} - \beta_u(u,v)
\frac{\partial}{\partial u} - \beta_v(u,v)
\frac{\partial}{\partial v} - \frac{\mathcal N}2 \eta_\psi(u,v)  \right. \nn \\
&& \ \ \ \ \ \ \left.
 - \gamma(u,v)  m \frac{\partial}{\partial m}\right]
  {\Gamma}^{(\mathcal N)}(k_i; m ,u, v, \mu)=0,  \label{eq:RG1}
\end{eqnarray}
where we have introduced the scaling functions
\begin{eqnarray}
&&\beta_u(u,v)= -\left.\mu\frac{\partial u}{\partial \mu} \right|_{0},  \ \ \ \
  \beta_v(u,v)= -\left.\mu\frac{\partial v}{\partial \mu} \right|_{0}, \label{beta}  \ \ \ \ \ \\
&&\eta_\psi(u,v)= - \beta_u(u,v)\frac{\partial \ln Z_\psi}{\partial u}- \beta_v(u,v)\frac{\partial \ln Z_\psi}{\partial v}, \ \ \ \  \\
&&\eta_m(u,v)= - \beta_u(u,v)\frac{\partial \ln Z_m}{\partial u} - \beta_v(u,v)\frac{\partial \ln Z_m}{\partial v}, \ \ \ \ \ \\
&& \gamma(u,v) = \eta_m(u,v)- \eta_\psi(u,v). \ \ \ \ \label{gam}
\end{eqnarray}
The subscript "0" stands for derivatives at fixed $u_0$, $v_0$ and $m_0$.
The dimensional analysis gives
\begin{eqnarray}
&&\!\!\!\!\!\!\!\!\!\! \Gamma^{(\mathcal N)}(k_i; m, u, v, \mu) =\lambda^{-d+{\mathcal N}(d-1)/2}\nonumber\\
&&\ \ \ \ \ \ \ \ \times {\Gamma}^{(\mathcal N)}(\lambda k_i; \lambda m, u, v, \lambda \mu),
\end{eqnarray}
which can be rewritten in an infinitesimal form  as
\begin{eqnarray}
&&\left[\mu\frac{\partial}{\partial \mu} + \sum\limits_i k_i \frac{\partial}{\partial k_i}
+ m \frac{\partial}{\partial m} \right. \nn \\
&&\ \ \ \ \ \ \left. -d+\frac{{\mathcal N}(d-1)}2 \right]
{\Gamma}^{(\mathcal N)}(k_i; m, u, v, \mu)=0. \ \ \ \  \label{eq:RG2}
\end{eqnarray}
Subtracting Eq.~(\ref{eq:RG1}) from Eq.~(\ref{eq:RG2}) we arrive at
\begin{eqnarray}
&&\!\!\!\!\!\!\!\left[  \beta_u
\frac{\partial}{\partial u} + \beta_v
\frac{\partial}{\partial v} + \sum\limits_i k_i \frac{\partial}{\partial k_i} + (1+\gamma ) m  \frac{\partial}{\partial m}   \right. \nn \\
&& \   \left.
  + \frac{\mathcal N}2 \left[ d-1   + \eta_\psi \right]-d \right] {\Gamma}^{(\mathcal N)}(k_i; m, u, v)=0. \ \ \label{eq:RG3}
\end{eqnarray}
Equation~(\ref{eq:RG3}) is a linear first order partial differential equation which can be solved by the method of
characteristics~\cite{Amit}. It reduces this equation to a set of ordinary differential equations which determines a family of curves
along which the solution can be integrated from some initial conditions given on a suitable hypersurface.
The characteristics lines of Eq.~(\ref{eq:RG3}) can be found from the flow equations
\begin{eqnarray}
&&   \frac{d \tilde{u} (\xi)}{d \ln \xi} = \beta_u(\tilde{u}(\xi),\tilde{v}(\xi)), \label{eq:fl1-1} \\
&&   \frac{d \tilde{v} (\xi)}{d \ln \xi} = \beta_v(\tilde{u}(\xi),\tilde{v}(\xi)), \\
&&   \frac{d \tilde{k}_i (\xi)}{d \ln \xi} = \tilde{k}_i(\xi), \\
&&   \frac{d \tilde{m} (\xi)}{d \ln \xi} = [1+\gamma(\tilde{u}(\xi), \tilde{v}(\xi))] \tilde{m}(\xi)  \label{eq:fl1-L}
\end{eqnarray}
with the initial conditions $\tilde{u} (1)=u$, $\tilde{v} (1)=v$, $\tilde{k}_i(1)=k_i$, $\tilde{m} (1)=m$.  The solution
of Eq.~(\ref{eq:RG3}) propagates along the characteristic curves (\ref{eq:fl1-1})-(\ref{eq:fl1-L}) according to
\begin{eqnarray}
&&   \frac{d \ln M_{\mathcal N}(\xi)}{d \ln \xi} = d - \frac{\mathcal N}2 [d-1 + \eta_\psi(\tilde{u}(\xi), \tilde{v}(\xi))],
\end{eqnarray}
with the initial condition $M_{\mathcal N}(1)=1$, and thus, satisfies the scaling relation
\begin{eqnarray}
  {\Gamma}^{(\mathcal N)}( \tilde{k}_i(\xi); \tilde{m}(\xi), \tilde{u}(\xi), \tilde{v}(\xi))=
M_{\mathcal N} (\xi) {\Gamma}^{(\mathcal N)}(k_i; m, u, v ). \nn \\ \label{eq:sol-rg1} \ \ \
\end{eqnarray}
$M_{\mathcal N} (\xi)$ encodes the anomalous scaling dimension of the fields $\psi$ and $\bar{\psi}$.
The critical behavior of the system is expected to be controlled by a stable fixed point (FP)
of the RG flow which is defined as simultaneous zero of $\beta$-functions~(\ref{beta}):
\begin{eqnarray}
\beta_u(u^*,v^*)=0, \ \ \ \beta_v(u^*,v^*)=0. \label{eq:fp-def}
\end{eqnarray}
Stability of a FP can be determined from the eigenvalues of the stability matrix
\begin{equation}\label{smatrix}
\mathcal{M}=\left(\begin{array}{c c}
\frac{\partial \beta_u(u,v)}{\partial u} & \frac{\partial \beta_u(u,v)}{\partial v}\\
\frac{\partial \beta_v(u,v)}{\partial u} &\frac{\partial \beta_v(u,v)}{\partial v}\end{array}\right).
\end{equation}
The FP is stable provided that both eigenvalues calculated at the FP~(\ref{eq:fp-def}) have negative real parts.
We can identify $\xi$  in Eqs.~(\ref{eq:fl1-1})-(\ref{eq:sol-rg1})
with the  correlation length. Then the solution~(\ref{eq:sol-rg1}) can be written in the vicinity of the FP~(\ref{eq:fp-def}) as
\begin{eqnarray}
 {\Gamma}^{(\mathcal N)}(k_i, m)= \xi^{{\mathcal N} d_\psi - d } f_{\mathcal N}( k_i \xi, m \xi^{1/\nu}), \ \ \
\end{eqnarray}
where we have identified the correlation length exponent
 \begin{eqnarray} \label{nufp}
\frac1{\nu} = 1+\gamma(u^*,v^*),
\end{eqnarray}
and the anomalous scaling dimension of the fields $\psi$ and $\bar{\psi}$
\begin{eqnarray}
d_\psi = \frac12[d-1+\eta_\psi(u^*,v^*)].
\end{eqnarray}
For instance, in the critical point we have
\begin{eqnarray}
 \overline{ \left\langle\bar{\psi}(r) \psi(0)\right\rangle } \sim {r^{-2d_\psi}}.
\end{eqnarray}

\section{\label{IV} Fixed points, their stability and scaling behavior}
\subsection{Renormalization to two-loop order }

In order to renormalize the theory~(\ref{eq:action-r}) to two-loop order we need the diagrams contributing to the two- and four-point vertex functions $\Gamma^{(2)}(p)$ and $\Gamma^{(4)}(p_i=0)$ in the replica limit $n\to 0$ which are shown in Fig.~\ref{fig:one-loop-diagrams} and Fig.~\ref{fig:two-loop-diagrams}.
In the one-loop approximation there are two diagrams contributing to the two-point vertex function $\Gamma^{(2)}(p)$ each of which we split into two parts
$A_1^{a}$ and $A_2^{a}$. The first part is computed at zero external momentum and the second part is the part which is linear
in the external momentum $\vec{p}$. The same is applied to the two-loop diagrams $C_2^{ab}$ and $C_3^{ab}$.
The diagrams contributing to the four-point function are computed at zero external momenta and expanded in small parameters  $\varepsilon$ and $\delta$
keeping the ratio $\frac{ \varepsilon }{\delta }$ finite.
Within the minimal subtraction scheme  we need only the poles in $\varepsilon$ and $\delta$ for the two-loop diagrams while for
the one-loop diagrams  one has to keep also the contributions  which are finite in the limit $\varepsilon,\delta\to 0$.
The poles of the diagrams shown in Fig.~\ref{fig:one-loop-diagrams} and Fig.~\ref{fig:two-loop-diagrams} are calculated with the help of the
formulas given in Appendix~\ref{loop-integrals}  and collected in Tables~\ref{table0} and \ref{table1}, respectively.
The vertex functions $\Gamma^{(2)}(p)$, $\Gamma_u^{(4)}(p_i=0)$ and $\Gamma_v^{(4)}(p_i=0)$  are computed in
Appendix~\ref{appendix:vertex-function}.
Using these functions  we find the $Z$-factors:
\begin{eqnarray}\label{Zu}
Z_u&=&1+ \frac{4u}{\varepsilon} +  \frac{4v}{\delta}
 -u^2 \left(\frac{2}{\varepsilon }-\frac{16}{\varepsilon ^2}\right) +\frac{4 v^3}{u(\varepsilon -3 \delta) } \nn\\
 &-& u v \left(\frac{28}{\delta +\varepsilon }-\frac{32}{\delta  \varepsilon }-\frac{8}{\delta
   }\right)- v^2 \left(\frac{10}{\delta }-\frac{4 \varepsilon }{\delta ^2}-\frac{16}{\delta^2}\right), \ \ \ \ \ \  \\
Z_v&=&1+  \frac{4u}{\varepsilon} +   \frac{4v}{\delta}
+ \frac{16 u^2 }{\varepsilon ^2} -\frac{4v^2}{\delta }  \left(1-\frac{ \varepsilon}{\delta}-\frac{4}{\delta}\right) \nn \\
&-& \frac{8 u v}{\delta +\varepsilon}  \left(1-\frac{ \varepsilon }{\delta}-\frac{4}{\delta  }-\frac{4}{\varepsilon }\right), \\
Z_m &=& 1 +    \frac{2u}{\varepsilon} +  \frac{2v}{\delta }
+\frac{6 u^2}{\varepsilon ^2} +v^2 \left(\frac{2
   \varepsilon }{\delta ^2}+\frac{6}{\delta ^2}-\frac{2}{\delta }\right)
\nn \\
&+& 4 u v \left(\frac{3}{\delta  \varepsilon }
 +\frac{1}{\delta}-\frac{2}{\delta +\varepsilon }\right)
, \\
\label{Zp}  Z_{\psi} &=& 1+ \frac{u^2}{\varepsilon }+\frac{4 u v \varepsilon }{\delta  (\delta +\varepsilon )}+\frac{v^2 (2
   \varepsilon -\delta )}{\delta ^2}.
\end{eqnarray}
For the SR disorder it was argued that the contribution coming from
the non-zero mass  in the numerator of the bare
propagator~(\ref{eq:bare-propagator}) vanishes, so that one can
neglect it from the beginning~\cite{schuessler09}. We have found
that this holds also for the case of the LR disorder at least to the
two-loop order, i.e. the contributions in the angular brackets in
Tables~\ref{table0} and \ref{table1} cancel each other in
Eqs.~(\ref{Zu})-(\ref{Zp}).
From Eqs.~(\ref{Zu})-(\ref{Zp}) using the definitions~(\ref{beta}) - (\ref{gam})  we obtain the
two-loop expressions for the $\beta$-functions
\begin{eqnarray}\label{bu}
\beta_u(u,v)&{=}& \varepsilon u {- }4u(  u {+} v) {+}8 u(u{+}v)^2
{+} 4v(u{+}v)^2 ,  \ \ \ \\
\beta_v(u,v) &{=}&  \delta v {-} 4v(  u {+} v) {+}4 v(u{+}v)^2, \ \ \ \label{bv}
\end{eqnarray}
 and for the other scaling functions giving the critical exponents:
\begin{eqnarray}
&&\eta_{\psi}(u,v)=-2 u^2+2 v^2 -\frac{4  \varepsilon }{\delta }u v-\frac{4  \varepsilon }{\delta }v^2,  \ \ \label{eq:eta-psi-exp} \\
&& \eta_{m}(u,v) = -2 u -2 v +4 u v +4 v^2 \nonumber\\
&& \ \ \ \ \ \ \ \ \ \ \ \ \ \ -\frac{4  \varepsilon }{\delta }u v-\frac{4  \varepsilon }{\delta }v^2, \nn
   \label{eq:eta-m-exp}  \\
&& \gamma(u,v) = -2(u+v) +2(u+v)^2. \label{gamma}
\end{eqnarray}
Note that the ratio $\frac{ \varepsilon }{\delta }$ is finite within our regularization scheme.
Though it is present in  the scaling functions $\eta_m$ and $\eta_\psi$,
all these ratios magically cancel each other in the $\beta$- and $\gamma$ - functions, leaving their coefficients pure integer constants.

\subsection{Expansions in small $\varepsilon$ and $\delta$}

We now analyze the renormalization group flow using expansion in small $\varepsilon$ and $\delta$. The $\beta$-functions have three FPs:
Gaussian, short-range correlated, and  long-range correlated disordered FPs.

\textit{(i) Gaussian fixed point}, given by
\begin{eqnarray}
u_{G}^*=  v_{G}^*=0,
\end{eqnarray}
describes the pure 2D Ising model with
the correlation length exponent $\nu_{\rm pure}=1/(1+\gamma(u^*_G,v^*_G))=1$.
Following  Ref.~\cite{shankar87} one  can  estimate singularity in the free energy.
Using the action~(\ref{eq:action-Dirac})
one can express the partition function of the Ising model as $Z_{\rm Ising}^2 =\int D\bar{\psi} D\psi e^{-S_D} \sim \det\left[\slashed{\partial} +  m_0 \right] $ with $m_0 \equiv \tau$. Applying the identity $\ln \det = \mathrm{tr} \ln$ we find $F_{\rm sing} \sim \tau^2\ln \tau$,
so that  $C \sim \ln (\tau^{-1})$ and $\alpha_{\rm pure}=0$.

\textit{(ii) Short-range correlated disordered fixed point (SR FP)}, given by
\begin{eqnarray}
u_{SR}^*=\frac{\varepsilon}{4}+\frac{\varepsilon^2}{8}, \ \
v_{SR}^*= 0,
\end{eqnarray}
merges with the Gaussian FP
at $d=2$. This implies that the SR correlated disorder is marginally irrelevant in two dimensions. As a consequence
it results only in logarithmic corrections to the scaling behavior of the pure 2D Ising model.
The two-loop logarithmic corrections are calculated in Appendix~\ref{sec:corrections}.
For the correlation length and the specific heat we find
\begin{eqnarray} \label{eq:log-correct-cor-fun}
&&  \xi \sim \tau^{-1} (\ln \tau^{-1})^{1/2} \left [1+ o \left (\frac{\ln\ln \tau^{-1}}{\ln \tau^{-1}} \right) \right], \\
&&  C_\mathrm{sing} \sim \ln\ln \tau^{-1}  \left [1+ o \left (\frac{1}{\ln \tau^{-1}} \right) \right];
\label{eq:log-specific-heat}
\end{eqnarray}
\textit{i.e.}, the subdominant two-loop logarithmic corrections identically  vanish.

The Gaussian FP becomes unstable with respect to the LR correlated disorder
for $\delta>0$.
This reproduces the extended Harris criterion
\cite{weinrib-83}, which states that the critical behavior of the pure system
is modified by the LR correlated disorder  if $\nu_{\rm pure}<2/a$.
Indeed, substituting into the last relation $\nu_{\rm pure}=1$
one arrives at $a<2$ which means $\delta>0$.

\begin{figure}
\includegraphics[width=75mm]{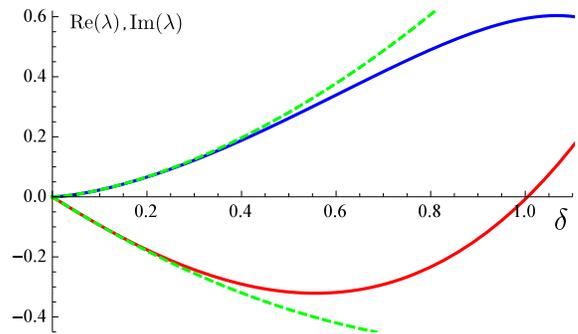}
\caption{The eigenvalues of the stability matrix in two dimensions ($d=2$)
as a function of $\delta$. There are two complex conjugated eigenvalues: the
red solid curve at the bottom is the real part and the blue curve at the top is the imaginary part.
The dashed lines are the series expansions (\ref{eq:eigenvalues-exp}).    }
  \label{fig:eigenvalues}
\end{figure}

\textit{(iii) Long-range correlated disordered fixed point (LR FP)} reads
\begin{eqnarray}
u_{LR}^*=\frac{\delta ^3}{16 (\delta -\varepsilon )}, \ \
v_{LR}^*= \frac{\delta }{4} -\frac{\delta^2 \varepsilon }{16 (\delta -\varepsilon )}.
\end{eqnarray}
In two dimensions  the LR  FP reduces to
\begin{eqnarray}\label{LRd2}
u_{LR}^*= \frac{\delta ^2}{16} + O(\delta^3),\ \ \ v_{LR}^*=\frac{\delta }{4} + O(\delta^3).
\end{eqnarray}
Let us perform the stability analysis of the LR FP.
The  two eigenvalues of the stability matrix~(\ref{smatrix}) computed at the LR FP~(\ref{LRd2})
at $d=2$ are shown in Fig.~\ref{fig:eigenvalues} as functions of $\delta$.
Both eigenvalues are complex
conjugated with the negative real parts for
$0<\delta< \delta_{\rm max}$, where the LR FP is stable. There are
no stable FPs for $\delta>\delta_{\rm max}$. Expansion of the
eigenvalues in small $\delta$ gives
\begin{eqnarray} \label{eq:eigenvalues-exp}
\lambda_{1,2}^{(LR)} &=& -\delta+\frac{\delta^2}2 + O(\delta^3) \pm \nonumber\\
&&\pm i \sqrt{\frac{\delta}2}
\left( \delta+\frac{\delta^2}4 + O(\delta^3) \right).
\end{eqnarray}
It is straightforward to see that
the value of $\delta_{\rm max}$ that follows from the expansion
(\ref{eq:eigenvalues-exp}) is $\delta_{\rm max}=2$, whereas
numerical diagonalization of the stability matrix~(\ref{smatrix})
gives $\delta_{\rm max} \approx 1.005$ (see
Fig.~\ref{fig:eigenvalues} for more details).
It is tempting to make more
precise the value of $\delta_{\rm max}$ by applying the familiar
resummation technique \cite{holovatch-02} to the two-loop series
(\ref{bu}), (\ref{bv}) at fixed $\epsilon$, $\delta$
\cite{Schloms,Blavatska2001}.  However, at $d=2$ (i.e. at
$\varepsilon=0$) the leading contribution to the first
$\beta$-function~(\ref{bu}) vanishes, making the series too short to
allow for a reliable resummation.

Thus, while according to the extended Harris criterion the LR FP may be stable
for $\delta>0$ we reveal  the existence  of the upper bound $\delta_{\rm max}$ for
its stability. Indeed, reasonable values of  $\delta$ lie between $0$ and $2$,
but $\delta=1$ corresponds to
the case of defect lines with random
orientation~\cite{Fedorenko2006}. One can argue that these lines may
break the 2D system into  disconnected domains: this is the argument
that can also be applied to the McCoy and Wu  model~\cite{McCoy68}.
Therefore one should take
values $\delta> 1$ with caution since strong
correlations may destabilize the LR FP and  drastically modify the
critical behavior. Since we cannot identify
any stable and perturbative in disorder FP for $\delta>\delta_{\rm max}$, two scenarios are possible:
(a) smearing of the sharp transition that is
manifested  in a runway of the renormalization group flow; (b) a new universality class
controlled by a non-perturbative infinite-randomness FP.
In the latter case one may expect relevance of rare regions which make
a difference between the typical and average correlations:
the correlation function between two arbitrary spins
separated by a large distance $x$ acquires a broad distribution~\cite{Fisher1992}.
Thus,
the typical correlation function is very different from the averaged one
which is dominated by rare strongly coupled regions of spins with atypical large
correlations. As a result,
there can be two correlations lengths, typical and averaged, and therefore
two critical exponents $\nu_{\rm typ}\le \nu_{\rm avr}$.

Substituting FP~(\ref{LRd2}) into  Eqs.~(\ref{nufp}) and (\ref{gamma}) we get
the correlation length exponent
\begin{eqnarray}
\frac1{\nu}=  1 -\frac{\delta}2 + O(\delta^3),
\end{eqnarray}
where the corrections of the second order in $\delta$ magically cancel each other.
Indeed, comparing Eq.~(\ref{bv}) and Eq.~(\ref{gamma}) one can observe that at least to two-loop order
\begin{equation}\label{betgam}
\beta_v(u,v)=v(\delta+2\gamma(u,v)).
\end{equation}
Calculating it at any FP and taking into account Eq.~(\ref{nufp}) we obtain
\begin{equation}
v^*(\delta+2(\nu^{-1}-1))=0, \label{eq:cong1}
\end{equation}
which  is in agreement with the conjecture
of Refs.~\cite{weinrib-83,HonkonenNalimov} that the identity
\begin{equation}\label{exact}
\nu=2/(2-\delta)=2/a
\end{equation}
is exact at the LR FP with $v^*\ne 0$.

\section{\label{V} Spin-spin correlation at criticality: bosonization}

We now focus on the scaling behavior of the two-point correlation function at criticality. Let us denote the correlation function
 in a given realization of disorder by $G(r)$ and introduce the set of critical exponents
\begin{eqnarray}
\overline{G(r)^N} \sim r^{-\eta_N}.
\end{eqnarray}
In the absence of multifractality  one expects $\eta_N=N \eta_1$ and
$\eta_1\equiv \eta$ is the standard pair correlation function
exponent. Since the correspondence  between the spin operators in
the Ising model and the Majorana fermions is non-local,
reexpressing the
 spin-spin correlation function in terms of fermions is complicated and is well defined only in two dimensions.  Thus the anomalous dimension calculated in Ref.~\cite{rajabpour08} from the scaling of the two-point fermionic correlation  function can not be directly
 connected with the critical exponent $\eta$.
Nevertheless using the Dirac representation allows one
to derive a compact formula for the square of the  correlation function \cite{shankar87}
\begin{eqnarray} \label{eq:spin-spin-cor-ferm}
{G(r)^2} = \left\langle \exp \left[ i\pi \int\limits_0^{r} dr' \bar{\psi}(r') \psi(r') \right] \right\rangle,
\end{eqnarray}
where the averaging is performed with the Dirac action~(\ref{eq:action-Dirac}). The
direct calculation of the spin-spin correlation function from the
fermionic representation (\ref{eq:spin-spin-cor-ferm}) has been
performed only for the pure system and involves a cumbersome algebra~\cite{dotsenko83}. A more simple way to get access to the spin-spin
correlation function is to use bosonization.  The latter maps the 2D
Dirac fermions (\ref{eq:action-Dirac}) into the sine-Gordon theory~\cite{shankar87,zinn-justin-book}
\begin{eqnarray} \label{eq:action-sine-Gordon-1}
S_{SG} = \int d^2 r \left\{ \frac12 (\nabla \varphi(r))^2 - \frac{\Lambda m(r)}{\pi} \cos\left[ \sqrt{4\pi} \varphi(r) \right] \right\},\nonumber\\
\label{sine}\end{eqnarray}
where $\Lambda$ is the UV cutoff.
The two-point spin correlation function becomes a two-point correlation function of the operator
\begin{eqnarray} \label{eq-operator-O}
\mathcal{O}(r) = \sin \sqrt{\pi} \varphi(r).
\end{eqnarray}
Note that  we bosonize the Dirac fermions so that this method gives not the two-point function but the square of the two-point function
\begin{eqnarray} \label{sq-two-point}
G(r)^2  =  \left\langle \mathcal{O}(r) \mathcal{O}(0)\right\rangle_{SG}
\end{eqnarray}
since the two Majorana fermions, i.e. two copies of the Ising model, have been combined to the Dirac fermions.
Averaging in (\ref{sq-two-point}) is performed with action (\ref{sine}). After averaging over disorder we obtain
\begin{eqnarray} \label{av-sq-two-point}
\overline{G(r)^2 } =  \overline{\left\langle \mathcal{O}(r) \mathcal{O}(0)\right\rangle_{SG}}.
\end{eqnarray}
To get a perturbative expansion for the correlation functions of the operator (\ref{eq-operator-O}) one
has to compute the correlation functions of exponentials of field $\varphi(r)$:
\begin{eqnarray}
\left\langle \prod_{j =1}^{n} e^{i \beta_j \varphi(r_j)} \right\rangle_0 &=& \int D\varphi
\exp[ - \frac12 \int d^2 r    (\nabla \varphi(r))^2  \nn \\
&&  + \sum_{j =1}^{n} i \beta_j \varphi(r_j)]. \label{eq:oo-cor1}
\end{eqnarray}
It can be shown~\cite{zinn-justin-book} that this correlation function is not vanishing only for
$\sum _{j =1}^{n} \beta_j=0$ and is given by
\begin{eqnarray}
\left\langle \prod_{j =1}^{n} e^{i \beta_j \varphi(r_j)} \right\rangle_0 =  \prod_{j<k}
( \Lambda|r_j-r_k|)^{\beta_j\beta_k/(2\pi)}. \label{eq:oo-cor2}
\end{eqnarray}
For the pure 2D Ising model at criticality, i.e  at $m(r)=0$, one finds
\begin{eqnarray}
G(r)^2 &=& \left\langle \mathcal{O}(r) \mathcal{O}(0)\right\rangle_0
 =  \frac{1}{2} \left\langle e^{i \sqrt{\pi} (\varphi(r)-\varphi(0))}   \right\rangle  \nn \\
 && = \frac12 (\Lambda r)^{-1/2},
\end{eqnarray}
and thus $\eta_{\mathrm{pure}}=1/4$ for the pure system.
We now calculate the first order correction in disorder.
Applying the replica trick to the action (\ref{eq:action-sine-Gordon-1}) we derive the
replicated action
\begin{eqnarray} \label{eq:action-sine-Gordon-2}
&& S = \frac12 \sum_{\alpha=1}^{n} \int d^2 r
(\nabla \varphi_\alpha(r))^2  - \frac{\Lambda^2}{2\pi^2} \sum_{\alpha,\beta=1}^{n} \nn \\
&&
  \int d^2 r d^2 r' g(r-r')
 \cos\left[ \sqrt{4\pi} \varphi_\alpha(r) \right] \cos\left[ \sqrt{4\pi} \varphi_\beta(r') \right]. \nn \\
\end{eqnarray}
Here we perform calculations directly in two dimensions to one-loop order
that allows us to put $u_0=0$. We calculate the averaged squared
spin-spin correlation function for one replica $\alpha=1$ to the
first order in $u_0$ and $v_0$ in Appendix~\ref{sec:correlation-function} and
obtain
\begin{eqnarray} \label{eq:two-point-6}
&& \left\langle \mathcal{O}_1(r) \mathcal{O}_1(0) \right\rangle_{S}  = \frac12 (\Lambda |r|)^{-1/2}
  \left[ 1 + \frac{u_0 \ln r\Lambda}{4\pi}  + \frac{v_0 |r|^{\delta}}{4 \pi \delta}
\right]. \nn \\
\end{eqnarray}
To renormalize the spin-spin correlation function we introduce the renormalization constant
\begin{eqnarray} \label{eq:two-point-7}
\mathring{\mathcal O}= Z_{\mathcal O}^{1/2} \mathcal{O}
\end{eqnarray}  which can be found from the relation
\begin{eqnarray} \label{eq:two-point-71}
\mathring{\overline{{G}(r)^2}} = Z_\mathcal{O}\, \overline{G(r)^2}.
\end{eqnarray}
Using the dimensional method developed in Ref.~\cite{Zoran2013} we can convert the logarithm in Eq.~(\ref{eq:two-point-6}) into a pole
as $\ln{r \Lambda} \to \frac{|r|^{\varepsilon}}{\varepsilon}$. Taking into account that
to the lowest order $u_0 = 2 m^{\varepsilon} v /{K_d} =4\pi m^{\varepsilon} u$ and
$v_0 = 2 m^{\delta} v /{K_d} =4\pi m^{\delta} v$
we obtain
\begin{eqnarray} \label{eq:two-point-8}
Z_{\mathcal O} = 1 + \frac{u}{\varepsilon}+ \frac{v}{\delta} + O(u^2, v^2).
\end{eqnarray}
The $\beta$-functions and the FP coordinates can be taken from the results
obtained for the Dirac fermions [Eqs.~(\ref{bu}), (\ref{bv}) and (\ref{LRd2})].
The resulting scaling function reads
\begin{eqnarray} \label{eq:two-point-81}
\eta_2 =  \frac12 - \beta_u \frac{\partial \ln Z_{\mathcal O}}{\partial u} - \beta_v \frac{\partial \ln Z_{\mathcal O}}{\partial v}
\end{eqnarray}
and to one-loop order is given by
\begin{eqnarray} \label{eq:two-point-91}
\eta_2 =  \frac12  - u  - v + O(u^2,v^2).
\end{eqnarray}
Using that to the one-loop order $u^*=0$ and $v^*=\delta/4$ [see Eq.~(\ref{LRd2})] we obtain
the critical exponent
\begin{eqnarray} \label{eq:two-point-9}
\eta_2 = \frac12 - \frac{\delta}4,
\end{eqnarray}
which describes algebraic decay of the square of the spin-spin correlation function  averaged over disorder:
\begin{eqnarray} \label{eq:two-point-10}
\overline{G(r)^2} = r^{-\eta_2}.
\end{eqnarray}
 Since $\overline{G^2} \ge \overline{G}^2$ and $\eta< \eta_{\rm pure}$ the exponent $\eta$ should satisfy the inequality
\begin{eqnarray} \label{eq:two-point-101}
\frac{\eta_2 }2 \approx  \frac14 - \frac{\delta}8 \le \eta \le  \frac14.
\end{eqnarray}
To go beyond the one-loop approximation is a nontrivial task  which is left
for a forthcoming study.

\section{\label{VI} Conclusions}

We have studied the 2D Ising model with LR correlated disorder using the mapping of the model to the
2D Dirac fermions in the presence of LR correlated random mass disorder. Using dimensional regularization
with double expansion in $\varepsilon=2-d$ and $\delta=2-a$  we renormalize the corresponding  field theory
up to the two-loop order. In two dimensions we have found two FPs: Gaussian FP [$u^*=0$, $v^*=0$] and LR FP
[$u^*=O(\delta^2)$, $v^*=O(\delta)$]. The Gaussian FP describes the 2D Ising model with SR disorder. The SR disorder
is marginally irrelevant in 2D  and leads to logarithmic correction to scaling. The SR FP is stable for  $\delta<0$
in accordance with the generalized Harris criterion $a \nu_{pure}  - d >0$ since   $\nu_{pure}=1$ in two dimensions.

We have shown that the LR FP is stable for $0<\delta<\delta_{max}$ with $\delta_{max}\approx 1.005$ to two-loop order. The LR FP
is characterized by the critical exponent $\nu=2/a + O(\delta^3)$ in accordance with the prediction $\nu=2/a $. Using mapping to the
sine-Gordon model we have also studied behavior of the  averaged square of the spin-spin correlation function at the LR FP
which has been found algebraically decaying with the distance as
$\overline{G^2(r)} \sim r^{-\eta_2}$. To the lowest order in disorder we have $ \eta_2 = \frac12 - \frac{\delta}4 + O(\delta^2)$ that gives the  bounds for the usual exponent $\eta$ describing the algebraic decay of the averaged correlation function:
$\frac12 \eta_2 \le \eta \le \frac14$.

We have not found a stable FP for $\delta>\delta_{max}$. This runaway  can be a sign of either a smeared  phase transition  or
a critical behavior controlled by an infinite randomness FP with different critical exponents. In the last case one can
expect  difference between the typical and averaged  correlation length exponents. The latter is supposed to be due to rare
regions with strong correlations so that one can expect $\nu_{\mathrm{avr}}>\nu_{\mathrm{typ}}$.  In order to study
the non-perturbative effects for $\delta>\delta_{\rm max}$ one can try to allow replica symmetry
breaking following  Refs.~\cite{dotsenko-95,fedorenko-01}.

Let us compare our finding with the known numerical results.
In Ref.~\cite{Bagamery2005} it was found that $\eta=0.2588(14)$ and $\nu=2.005(5)$ for  $a=1$ ($\delta=1$). The
exponent $\eta$ satisfies the inequality~(\ref{eq:two-point-101}) while the exponent $\nu$ is very close to the prediction  $\nu=2/a$.
In Ref.~\cite{Chatelain2014} it was found that   $\eta=0.204(14)$ and $\nu=7.14$ for $a=2/3$ ($\delta=4/3$).
It seems that the  exponent $\eta$ also satisfies the inequality but the exponent $\nu$ is much higher than the prediction corresponding
to the perturbative LR FP. That was ascribed to hyperscaling violation in the Griffiths phase due to large disorder fluctuations.
In the light of our work this is not surprising.
Indeed,  the runaway of the RG flow for $\delta>\delta_{\rm max}$ suggests that either
the system flows towards an inaccessible within a weak disorder RG infinite randomness FP  which controls the transition
or the transition is smeared out.  The numerical simulations of Ref.~\cite{Chatelain2014} are in favor of the first
scenario but this still remains an open question.

Another reason for such discrepancy may be due to peculiarities of the spatial distribution of disorder in the model
analyzed in~\cite{Chatelain2014}. There, the  spin configurations of
the Ashkin-Teller model at the critical point were used to construct correlated distribution of random couplings. In turn, these
displayed large self-similar clusters of strong/weak bonds \cite{Chatelain16}. Although, by construction, the disorder
correlations in Ref.~\cite{Chatelain2014} were governed by the power-law decay ($\ref{eq:dis-cor-0}$), formal description of their
impact might call for the model that differs from the one analyzed in our paper since the bare disorder distribution is
strongly non-Gaussian.
Note that all above values of the exponent $\nu$ satisfy the Chayes-Chayes-Fisher-Spencer inequality
for the correlation length exponent  of disordered systems, $\nu\ge 2/d$ \cite{Chayes1986}.
This indicates absence of difference between the intrinsic correlation length and the finite-size correlation lengths in this
problem.

\acknowledgments It is our pleasure to thank Christophe
Chatelain and Victor Dotsenko for stimulating discussions. MD and VB thank the Laboratoire
de Physique de l'ENS Lyon for hospitality during preparation of
this work.
This work was supported in part by the European Union's Research
and Innovation funding program FP7 IRSES Marie-Curie Grants No.
PIRSES-GA-2011-295302 ``Statistical Physics in Diverse Realizations'',
No. PIRSES-GA-2010-269139 ``Dynamics and Cooperative Phenomena in Complex
Physical and Biological Environments'' (MD and VB), and No. 612707
``Dynamics of and in Complex Systems'', No. 612669 ``Structure and
Evolution of Complex Systems with Applications in Physics and Life
Sciences''  (YuH).
AAF acknowledges support from the French Agence Nationale de la Recherche through
Grants No. ANR-12-BS04-0007 (SemiTopo), No. ANR-13-JS04-0005-01 (ArtiQ),
and No. ANR-14-ACHN-0031 (TopoDyn).

\appendix

\section{Vertex functions}  \label{appendix:vertex-function}
Here we present the  expressions for vertex functions  using diagrammatic presentation (see Fig.~\ref{fig:one-loop-diagrams} and  Fig.~\ref{fig:two-loop-diagrams}).
Taking into account the combinatorial factors for the diagrams, the two-point function is
\begin{eqnarray}\label{ver2}
\Gamma^{(2)}(p) &=& \sigma p \left\{ 1 + A_2^{1}v_0-C_{3}^{0,0} u_0^2 -(C_{3}^{1,0}+C_{3}^{0,1})u_0v_0 \right. \nn \\
&& \left. - C_{3}^{1,1} v_0^2 \right\} - i m_0 \left\{ 1 - A_1^0u_0 - A_1^1v_0  \right. \nn \\
&& + (C_{1}^{0,0} + C_{2}^{0,0})u_0^2 + (C_{1}^{1,0} + C_{1}^{0,1}+ C_{2}^{1,0} \nn \\
&& \left. + C_{2}^{0,1}) u_0v_0 + (C_{1}^{1,1}+C_{2}^{1,1})v_0^2 \right\}.
\end{eqnarray}
The SR part of the full four-point vertex functions reads:
\begin{eqnarray}\label{veru}
\Gamma_u^{(4)}(0) &=&u - 2B^{0,0}_1 u^2 - 2 B^{0,1}_1 uv + u^2 v (\tilde{D}^{1,0,0}_{12}+2 \tilde{D}^{1,0,0}_{14} \nn \\
&& +2 {D_1^{1,0,0}}+2 {D_2^{1,0,0}}+4 {D_3^{1,0,0}}+4 {D_4^{1,0,0}}+{D_5^{1,0,0}} \nn \\
&&+4 {\tilde{D}_6^{1,0,0}}+\tilde{D}_{12}^{0,1,0}+2 {\tilde{D}_{14}^{0,1,0}}+2 {D_1^{0,1,0}}+2 {D_2^{0,1,0}} \nn \\
&&+4 {D_3^{0,1,0}} +4 {D_4^{0,1,0}}+{D_5^{0,1,0}}+4 {\tilde{D}_8^{0,1,0}}+{\tilde{D}_{12}^{0,0,1}} \nn \\
&&  +2 {\tilde{D}_{14}^{0,0,1}}+4 {\tilde{D}_6^{0,0,1}})+u v^2 (\tilde{D}_{12}^{1,1,0}+2 \tilde{D}_{14}^{1,1,0} \nn \\
&&   +2 {D_1^{1,1,0}}+2 {D_2^{1,1,0}}+4 {D_3^{1,1,0}} +4{D_4^{1,1,0}} +{D_5^{1,1,0}} \nn \\
&&   +4 {\tilde{D}_6^{1,1,0}}+\tilde{D}_{12}^{1,0,1}  +2 \tilde{D}_{14}^{1,0,1}+4   {\tilde{D}_6}^{1,0,1}\nn \\
&&   +\tilde{D}_{12}^{0,1,1} +2 \tilde{D}_{14}^{0,1,1}+4   {\tilde{D}_6}^{0,1,1}) +v^3 (\tilde{D}_{12}^{1,1,1} \nn \\
&&    +2 \tilde{D}_{14}^{1,1,1}+4  {\tilde{D}_6^{1,1,1}}) +u^3 (\tilde{D}_{12}^{0,0,0}+2 \tilde{D}_{14}^{0,0,0}  \nn \\
&&    +2 {D_1^{0,0,0}} +2    {D_2^{0,0,0}}+4 {D_3^{0,0,0}} \nn \\
&& +4 {D_4^{0,0,0}}+{D_5^{0,0,0}}+4 {\tilde{D}_6^{0,0,0}}),
\end{eqnarray}
where $\tilde{D}^{a,b,c}_i=D^{a,b,c}_i+D^{a,b,c}_{i+1}$.  The LR part of the four-point vertex is given by
\begin{eqnarray}\label{verv}
\Gamma_v^{(4)}(0) &=&v- 2B^{1,1}_1 v^2 - 2 B^{1,0}_1 uv +v^3 (2 D_1^{1,1,1}+2
   D_2^{1,1,1}\nn \\
&&+4 {D_3^{1,1,1}}+4
  {D_4^{1,1,1}}+D_5^{1,1,1})+u v^2 (2
  D_1^{1,0,1}\nn \\
&&+2 {D_2^{1,0,1}}+4 {D_3^{1,0,1}}+4
   {D_4^{1,0,1}}+{D_5^{1,0,1}}+2 {D_1^{0,1,1}}\nn \\
&&+2
   D_2^{0,1,1}+4 {D_3^{0,1,1}}+4
   {D_4^{0,1,1}}+D_5^{0,1,1})\nn \\
&&+u^2 v (2 {D_1^{0,0,1}}+2 {D_2^{0,0,1}}+4{D_3^{0,0,1}}+4
   {D_4^{0,0,1}}\nn \\
&&+{D_5^{0,0,1}}).
\end{eqnarray}
The poles and finite
parts of the one-loop diagrams [$A^{a}_i$ in~(\ref{ver2}) and $B^{a,b}_i$
in~(\ref{veru})-(\ref{verv})  shown
in Fig.~\ref{fig:one-loop-diagrams} ] are given in  Table~\ref{table0}
together with their combinatorial factors. %
The poles of the two-loop diagrams [ $C^{a,b}_i$ in (\ref{ver2}) and
$D^{a,b,c}_i$ in (\ref{veru})-(\ref{verv}) shown in
Fig.~\ref{fig:two-loop-diagrams} ] are summarized in Table~\ref{table1}.
Some of the  two-loop integrals appearing in the calculations of
poles are summarized in  Appendix~\ref{loop-integrals}.
The angular brackets in
Tables~\ref{table0} and \ref{table1}
denote contributions resulting from the mass
in the numerator of the bare propagator~(\ref{eq:bare-propagator}).
These contributions cancel each other in the $Z$-factors~(\ref{Zu})-(\ref{Zp})
at least to two-loop order as  happens
in the case of uncorrelated disorder~\cite{schuessler09}.

\begin{table}[hptb]
\begin{tabular}{|c|c|c|}  \hline
 Diag. & Value & C.F. \\
\hline
 $A^{0}_{1}$ & $ \langle\frac{2}{\varepsilon}\rangle$  &1 \\
  $A^{1}_{1}$ & $ \langle\frac{2}{\delta}\rangle$ &1 \\
  $A^{0}_{2}$ & $0$ &1 \\
   $A^{1}_{2}$ & $1-\frac{\varepsilon}{\delta}$ &1 \\
\hline
 $B^{0,0}_{1}=B^{1,0}_{1}$ & $\frac{2}{\varepsilon}-1 - \langle1\rangle$ &2 \\
  $B^{0,1}_{1}=B^{1,1}_{1}$ & $\frac{2}{\delta}-1 - \langle1\rangle$ &2 \\
 $B^{a,b}_{2}+B^{a,b}_{3}$ & $ - \langle1\rangle $ &2 \\
 \hline
\end{tabular}
\caption{Poles and finite parts of one-loop diagrams in the units of {\protect $\frac{K_d}{2}$}. C.F. is the combinatorial
factor.  The angular brackets denote contribution resulting from the mass in the numerator of the bare
propagator~(\ref{eq:bare-propagator}).
  \label{table0}}
\end{table}

\begin{table*}[hptb]
\begin{tabular}{|c|c|c|c|}  \hline
Diagram & Different vertices & Poles & C.F. \\
 &$\{a,b,c\}$ & & \\
\hline

$C^{a,b}_1$ &$\{0,0\}$ & $\frac{4}{\varepsilon^2}\left(1-\varepsilon\right)$ & 1 \\
 &$\{1,0\}$  & $\frac{4}{\delta\varepsilon}\left(1-\delta\right)$ &1 \\
 &$\{0,1\}$ & $\frac{8}{\delta(\varepsilon+\delta)}-\frac{8\varepsilon}{\delta(\varepsilon+\delta)}$ &1 \\
 &$\{1,1\}$  & $\frac{2(3\delta-\varepsilon)}{\delta^2(2\delta-\varepsilon)}-\frac{2(\delta+\varepsilon)}{\delta^2}$ &1 \\
 \hline
$C^{a,b}_2$ &$\{0,0\}$  & $\frac{2}{\varepsilon^2}$ & 1\\
 & $\{1,0\}$,$\{0,1\}$  & $\frac{4}{\varepsilon(\delta+\varepsilon)}$& 1\\
 & $
 \{1,1\}$  & $\frac{2(3\delta-2\varepsilon)}{\delta^2(2\delta-\varepsilon)}$& 1\\
 \hline
 $C^{a,b}_3$ & $\{0,0\}$  & $ \frac1{\varepsilon}$ &1 \\
 & $\{1,0\}$,$\{0,1\}$  & $\frac{2}{\varepsilon}-\frac{2}{\varepsilon+\delta}$ & 1 \\
 & $\{1,1\}$ & $\frac{(3\delta-2\varepsilon)}{\delta^2}$ & 1 \\
 \hline
$D^{a,b,c}_1$ & $\{0,0,0\}$, $\{0,0,1\}$  & $\frac{4}{\varepsilon^2}\left(1-\varepsilon\right)-\left<\frac{4}{\varepsilon}\right>$&2\\
 & $\{1,0,0\}$, $\{1,0,1\}$ & $\frac{4}{\varepsilon\delta}\left(1-\frac{\varepsilon+\delta}{2}\right)-\left<\frac{2}{\varepsilon}\right>-\left<\frac{2}{\delta}\right>$&2\\
 & $\{0,1,0\}$, $\{0,1,1\}$ & $\frac{8}{\delta(\varepsilon+\delta)}\left(1-\frac{\varepsilon+\delta}{2}\right)-\left<\frac{4}{\delta}\right>$&2\\
 & $\{1,1,0\}$,$\{1,1,1\}$ & $\frac{2(3\delta-\varepsilon)}{\delta^2(2\delta-\varepsilon)}-\frac{2(3\delta-\varepsilon)}{\delta(2\delta-\varepsilon)}-
 \left<\frac{2(3\delta-\varepsilon)}{\delta(2\delta-\varepsilon)}\right>$&2\\
 \hline
$D^{a,b,c}_2$ & $\{0,0,0\}$,$\{0,0,1\}$ & $-\frac{2}{\varepsilon^2}+\frac{2}{\varepsilon}+\left\langle\frac{2}{\varepsilon}\right\rangle$&2\\
 & $\{1,0,0\}$,$\{0,1,0\}$,$\{1,0,1\}$,$\{0,1,1\}$ & $-\frac{4}{\delta(\varepsilon+\delta)}+\frac{2}{\delta}+\left\langle\frac{2}{\delta}\right\rangle$&2\\
  & $\{1,1,0\}$,$\{1,1,1\}$ & $-\frac{2}{\delta(2\delta-\varepsilon)}+\frac{2}{2\delta-\varepsilon}+\left\langle\frac{2}{2\delta-\varepsilon}\right\rangle$&2\\
 \hline
$D^{a,b,c}_3$ & $\{0,0,0\}$,$\{0,0,1\}$ & $\frac{2}{\varepsilon^2}-\frac{2}{\varepsilon}-\left<\frac{2}{\varepsilon}\right>$&4\\
 & $\{1,0,0\}$,$\{1,0,1\}$ & $\frac{4}{\varepsilon(\varepsilon+\delta)}-\frac{2}{\varepsilon}-\left<\frac{2}{\varepsilon}\right>$&4\\
 & $\{0,1,0\}$,$\{0,1,1\}$ & $\frac{4}{\delta(\varepsilon+\delta)}-\frac{2}{\delta}-\left<\frac{2}{\delta}\right>$&4\\
 & $\{1,1,0\}$,$\{1,1,1\}$ & $\frac{2}{\delta^2}-\frac{2}{\delta}-\left<\frac{2}{\delta}\right>$&4\\
 \hline
$D^{a,b,c}_4$ & $\{0,0,0\}$,$\{0,0,1\}$,$\{1,0,0\}$,$\{1,0,1\}$ & $-\left<\frac{2}{\varepsilon}\right>$&4\\
 & $\{0,1,0\}$,$\{0,1,1\}$ & $-\left<\frac{2}{\delta}\right>+\frac{2(\delta-\varepsilon)}{\delta(\varepsilon+\delta)}$&4\\
 & $\{1,1,0\}$,$\{1,1,1\}$ & $-\left<\frac{2}{\delta}\right>+\frac{(\delta-\varepsilon)}{\delta^2}$ &4 \\
 \hline
$D^{a,b,c}_5$ & $\{0,0,0\}$,$\{0,0,1\}$ & $\frac{4}{\varepsilon^2}-\frac{4}{\varepsilon}-\left\langle\frac{4}{\varepsilon}\right\rangle$&1\\
 & $\{1,0,0\}$,$\{0,1,0\}$,$\{1,0,1\}$,$\{0,1,1\}$ & $\frac{4}{\varepsilon\delta}-\frac{2}{\varepsilon}-\frac{2}{\delta}-\left\langle \left(\frac{2}{\varepsilon}+\frac{2}{\delta}\right)\right\rangle$&1\\
 & $\{1,1,0\}$,$\{1,1,1\}$ & $\frac{4}{\delta^2}-\frac{4}{\delta}-\left\langle\frac{4}{\delta}\right\rangle$&1\\
 \hline
$D^{a,b,c}_6+D^{a,b,c}_7$ & $\{0,0,0\}$,$\{0,1,0\}$,$\{0,0,1\}$,$\{0,1,1\}$  & $-\left\langle\frac{4}{\varepsilon}\right\rangle$ & 4\\
 & $\{1,0,0\}$,$\{1,1,0\}$,$\{1,0,1\}$,$\{1,1,1\}$ & $-\left\langle\frac{4}{\delta}\right\rangle$ & 4 \\
 \hline
$D^{a,b,c}_8+D^{a,b,c}_9$ & $\{a,b,c\}$ & $0$ &2\\
\hline
$D^{a,b,c}_{10}+D^{a,b,c}_{11}$ & $\{a,b,c\}$ & $0$ &2\\
 \hline
 $D^{a,b,c}_{12}+D^{a,b,c}_{13}$ & $\{0,0,0\}$ & $\frac{4}{\varepsilon^2}-\frac{2}{\varepsilon}$ &1 \\
 & $\{1,0,0\}$,$\{0,0,1\}$ & $\frac{4}{\delta\varepsilon}-\frac{2}{\varepsilon}$&1 \\
 & $\{0,1,0\}$ & $\frac{8}{\delta(\delta+\varepsilon)}-\frac{4}{\delta+\varepsilon}$ &1 \\
 & $\{1,1,0\}$,$\{0,1,1\}$ & $\frac{2(3\delta-\varepsilon)}{\delta^2(2\delta-\varepsilon)}-\frac{3\delta-\varepsilon}{\delta^2}$ &1 \\
 & $\{1,0,1\}$ & $\frac{4}{\delta^2}-\frac{2(2\delta-\varepsilon)}{\delta^2}$ &1 \\
 & $\{1,1,1\}$ & $\frac{8}{(2\delta-\varepsilon)(3\delta-\varepsilon)}-\frac{4(3\delta-2\varepsilon)}{(2\delta-\varepsilon)(3\delta-\varepsilon)}$ &1 \\
 \hline
 $D^{a,b,c}_{14}+D^{a,b,c}_{15}$   & $\{0,0,0\}$ & $-\frac{2}{\varepsilon^2}+\frac{2}{\varepsilon}$ &2 \\
 & $\{1,0,0\}$ & $\frac{4}{\delta(\delta+\varepsilon)}-\frac{4}{\varepsilon \delta}+\frac2{\varepsilon}$ &2 \\
 & $\{0,1,0\}$,$\{0,0,1\}$ & $-\frac{4}{\delta(\delta+\varepsilon)}+\frac4{\delta+\varepsilon}$ &2 \\
 & $\{1,1,0\}$,$\{1,0,1\}$ & $-\frac{2}{\delta^2}+\frac3{\delta}-\frac{\varepsilon}{\delta^2}$ &2 \\
 & $\{0,1,1\}$ & $-\frac{2}{\delta(2\delta-\varepsilon)}+\frac{2}{\delta}$ &2 \\
 & $\{1,1,1\}$ & $-\frac{4}{(2\delta-\varepsilon)(3\delta-\varepsilon)}-\frac{2}{2\delta-\varepsilon}+\frac{8}{3\delta-\varepsilon} $ &2 \\
 \hline
\end{tabular}

\caption{\label{table1} Poles of two-loop diagrams in the units of {\protect $\hat K$}.  C.F. is the combinatorial
factor. The angular brackets denote contribution resulting from the mass in the numerator of the bare
propagator~(\ref{eq:bare-propagator}). }

\end{table*}


\section{\label{loop-integrals} Table of two-loop integrals}
Here we provide the list of the two-loop integrals, which are  helpful in
calculation of the two-loop diagrams.
To calculate these integrals we used the methods based on the hypergeometric
function representation which were
developed in Ref.~\cite{Dudka15} for the $\varphi^4$ - model with correlated disorder.
We introduce the shortcut notations $[1]:=q_1^2+m^2$,
$[2]:=q_2^2+m^2$, $[3]:=(q_1+q_2)^2+m^2$,  $\hat K=\frac{K_d^2}{4}$ as well as
shortcut notation for the integration $\int=\int_{\vec{q_1}}\!\int_{\vec{q_2}}$.
Only the poles are shown so that the omitted terms are
of order $O(1)$ unless something else is explicitly stated.

\subsection{$a=b=c=0$}

\begin{equation}
\int\frac1{[1][2]}{=}\int\frac1{[1][3]}{=}\hat K m^{-2\varepsilon}
  \left[ \frac4{\varepsilon^2} \right],
\end{equation}

\begin{equation}
\int\frac1{[1][3]^2}{=}\int\frac1{[1][2]^2}{=}\hat K m^{-2\varepsilon-2}
  \left[ \frac2{\varepsilon} + O(\varepsilon) \right].
\end{equation}


\subsection{$a\ne 0$}

\begin{equation}
\int\frac{q_1^{a-d}}{[1][2]}{=}\int\frac{q_1^{a-d}}{[1][3]}{=}\hat K m^{-\varepsilon-\delta}
  \left[ \frac4{\varepsilon\delta} \right],
\end{equation}

\begin{equation}
\int\frac{q_1^{a-d}}{[2][3]}=\hat K m^{-\varepsilon-\delta}
  \left[ \frac8{\delta(\delta+\varepsilon)}  \right],
\end{equation}

\begin{equation}
\int\frac{q_1^{a-d}}{[1][2]^2}=\int\frac{q_1^{a-d}}{[1][3]^2}
=\hat K m^{-\varepsilon-\delta-2}
  \left[ \frac2{\delta} \right],
\end{equation}

\begin{equation}
\int\frac{q_1^{a-d}}{[2]^2[3]}=\int\frac{q_1^{a-d}}{[2][3]^2}
=\hat K m^{-\varepsilon-\delta-2}
  \left[ \frac2{\delta}  \right],
\end{equation}

\begin{equation}
\int\frac{q_1^2 q_1^{a-d}}{[2]^2[3]^2}
=\hat K m^{-\varepsilon-\delta-2}
  \left[ \frac4{\delta}  \right],
\end{equation}

\begin{equation}
\int\frac{ q_1^{a-d}}{[1]^2[2]}
=\hat K m^{-\varepsilon-\delta-2}
  \left[ \frac2{\varepsilon} \right],
\end{equation}

\begin{equation}
\int\frac{q_1^2 q_1^{a-d}}{[2]^3[3]}
=\hat K m^{-\varepsilon-\delta-2}
  \left[ - 1 + O(\delta, \varepsilon) \right],
\end{equation}

\begin{equation}
\int\frac{ q_1^{a-d}}{[2]^3[3]}
=\hat K m^{-\varepsilon-\delta-4}
  \left[ \frac1{\delta}  \right].
\end{equation}

\subsection{$a\ne 0,b\ne 0$}

\begin{equation}
\int\frac{q_1^{a-d}q_2^{a-d}}{[1][2]}
=\hat K m^{-2\delta}
  \left[ \frac4{\delta^2}  \right],
\end{equation}

\begin{eqnarray}
\int\frac{q_1^{a-d}q_2^{a-d}}{[1][3]}{=}\!\int\frac{q_1^{a-d}q_2^{a-d}}{[2][3]}{=}\hat K m^{{-}2\delta}
 \! \left[ \frac{2(3\delta{-}\varepsilon)}{\delta^2(2\delta{-}\varepsilon)}  \right],
\end{eqnarray}

\begin{eqnarray}
\int\frac{q_1^{a-d}q_2^{a-d}}{[1][3]^2}{=}\int\frac{q_1^{a-d}q_2^{a-d}}{[2][3]^2}
{=}\hat K m^{-2\delta-2}
  \left[ \frac{2}{2\delta{-}\varepsilon}  \right],
\end{eqnarray}

\begin{eqnarray}
\int\frac{q_1^{a-d}q_2^{a-d}}{[2]^2[3]}
 {=} \int\frac{q_1^{a-d}q_2^{a-d}}{[1][2]^2}
{=}\hat K m^{-2\delta-2}
  \left[ \frac{2}{\delta}  \right],
\end{eqnarray}

\begin{eqnarray}
&&\int\frac{q_2^2 q_1^{a-d}q_2^{a-d}}{[1]^2[3]}
=\hat {K} m^{-2\delta}
  \left[ \frac{2(3\delta-\varepsilon)}{\delta^2 (2\delta-\varepsilon)}
  - \frac{2(3\delta-\varepsilon)}{\delta^2} \right], \nonumber \\
\end{eqnarray}

\begin{eqnarray}
&&\int\frac{[2] q_1^{a-d}q_2^{a-d}}{[1]^2[3]^2}
{=}\hat K m^{-2\delta-2}
  \left[ \frac{2(3\delta-\varepsilon)}{\delta (2\delta-\varepsilon)} \right],
\end{eqnarray}

\begin{eqnarray}
\int\frac{[2] q_1^{a-d}q_2^{a-d}}{[1][3]^2}
{=}\hat K m^{-2\delta}
  \left[ \frac{2(3\delta-\varepsilon)}{\delta^2 (2\delta-\varepsilon)}
  - \frac{2}{\delta} \right],
\end{eqnarray}

\begin{eqnarray}
&&\int\frac{[2] q_1^{a-d}q_2^{a-d}}{[1]^2[3]}
{=}\hat K m^{-2\delta}
  \left[ \frac{2(3\delta-\varepsilon)}{\delta^2 (2\delta{-}\varepsilon)}{ -}
  \frac{2(2\delta{-}\varepsilon)}{\delta^2} \right],\nonumber\\
\end{eqnarray}

\begin{eqnarray}
\int\!\!\frac{q_2^4 q_1^{a{-}d}q_2^{a{-}d}}{[1]^2[3]^2}
{=}\hat K m^{{-}2\delta}
\!\! \left[ \frac{8(3\delta{-}\varepsilon)}{\delta^2 (2\delta{-}\varepsilon)}
{- } \frac{2(8\delta{-}3\varepsilon)(3\delta{-}\varepsilon)}{\delta^2(2\delta{-}
\varepsilon)} \right]. \nonumber\\
\end{eqnarray}

\subsection{$b\ne 0,c\ne 0$}

\begin{eqnarray}
\int\frac{q_2^{2(a-d)}}{[1][2]}=\int\frac{q_2^{2(a-d)}}{[2][3]}
=\hat K m^{-2\delta}
  \left[ \frac{4}{\varepsilon(2\delta{-}\varepsilon)} \right],
\end{eqnarray}

\begin{equation}
\int\frac{q_2^{2(a-d)}}{[1][3]}
=\hat K m^{-2\delta}
  \left[ \frac{4}{\delta(2\delta-\varepsilon)} \right],
\end{equation}

\begin{eqnarray}
\int\frac{q_2^{2(a-d)}}{[1][2]^2}&=&\int\frac{q_2^{2(a-d)}}{[2]^2[3]}=\hat K m^{-2\delta-2}
  \left[ \frac{2}{\varepsilon}  \right].
\end{eqnarray}

\subsection{$a\ne 0,b\ne 0,c\ne 0$}

\begin{eqnarray}
\int\frac{q_1^{a-d}q_2^{2(a-d)}}{[1][2]}
{=}\hat K m^{-3\delta+\varepsilon}
  \left[ \frac{4}{\delta(2\delta-\varepsilon)} \right],
\end{eqnarray}

\begin{eqnarray}
\int\!\frac{q_1^{a{-}d}q_2^{2(a{-}d)}}{[1][3]}
{=}\hat K m^{{-}3\delta{+}\varepsilon}
 \!\! \left[\! \frac{4(5\delta{-}3\varepsilon)}{(3\delta{-}2\varepsilon)
 (2\delta{-}\varepsilon)(3\delta{-}\varepsilon)}\!\right],
\end{eqnarray}

\begin{eqnarray}
\int\frac{q_1^{a{-}d}q_2^{2(a{-}d)}}{[2][3]}
{=}\hat K m^{{-}3\delta{+}\varepsilon}
  \left[
 \frac{8(2\delta{-}\varepsilon)}{\delta(3\delta{-}2\varepsilon)
 (3\delta{-}\varepsilon)}  \right],
\end{eqnarray}

\begin{equation}
\int\frac{q_1^{a-d}q_2^{2(a-d)}}{[2]^2[3]}
=\hat K m^{-3\delta+\varepsilon-2}
  \left[ \frac{2}{\delta}  \right],
\end{equation}

\begin{eqnarray}
\int\frac{q_1^{a{-}d}q_2^{a{-}d} |q_1{+}q_2|^{a{-}d}}{[1][2]}
{=}\hat K m^{{-}3\delta{+}\varepsilon}
  \left[ \frac{8}{(2\delta{-}\varepsilon)(3\delta{-}\varepsilon)}  \right].\nonumber\\
\end{eqnarray}

\section{Logarithmic corrections for SR disorder} \label{sec:corrections}

In order to calculate the subdominant logarithmic corrections to scaling behavior in two dimensions
due to SR disorder we have to find the asymptotic flow to the Gaussian FP.
Here we do this to two-loop order. The flow equations read
\begin{eqnarray}
&&   \frac{d u }{d l } = \beta_u(u,v=0) = -4 u^2 + 8 u^3 + O(u^4), \label{eq:cor1} \\
&&   \frac{d { \ln \tau }}{d l} = - [1+\gamma(u, 0)] = -1+2u-2u^2+O(u^3) ,  \label{eq:cor2} \ \ \ \ \ \ \ \  \\
&&   \frac{d \ln F}{d l} = \gamma(u, 0) = -2u+2u^2+O(u^3), \label{eq:cor3}
\end{eqnarray}
where $l=\ln \xi$ and $F$ is the vertex function with insertion of the composite operator $\bar{\psi} (0) \psi (0)$
defined in Refs.~\cite{jug96,Shalaev94}.
The asymptotic  behavior of the solution of Eq.~(\ref{eq:cor1}) in the limit $l \to \infty$ is
\begin{eqnarray}
&&   u(l)= \frac1{4l}+\frac{\ln l}{8 l^2} + O\left(\frac1{l^2}\right). \label{eq:log-cor-solution}
\end{eqnarray}
Substituting the flow (\ref{eq:log-cor-solution}) to Eq.~(\ref{eq:cor2})  we obtain
\begin{eqnarray}
&&   \tau^{-1} \sim \xi (\ln \xi)^{-1/2} \left [1+\frac{\ln\ln \xi}{4\ln\xi} \right].
\end{eqnarray}
Inverting this equality with logarithmic accuracy we arrive at Eq.~(\ref{eq:log-correct-cor-fun}).
The singular part of the specific heat in the asymptotic regime
is given by $C_{\mathrm{sing}} = \int dl F^2(l)$ \cite{jug96}. Solving Eq.~(\ref{eq:cor3}) we obtain
\begin{eqnarray}
&&   C_{\mathrm{sing}}(l) = \ln l  \left[1 - \frac1{2\ln l }\right].
\end{eqnarray}
Using $l=\ln \xi$ where $\xi$ is given by Eq.~(\ref{eq:log-correct-cor-fun}) we derive Eq.~(\ref{eq:log-specific-heat}).

\section{Correlation function} \label{sec:correlation-function}
We now calculate the  two-point
function~(\ref{eq:two-point-6}) for the replica $\alpha=1$
to the lowest order in disorder.
The first-order correction in disorder  can be
split into the SR and LR parts as follows:
\begin{eqnarray} \label{eq:two-point-0}
&&\left\langle \mathcal{O}_1(r) \mathcal{O}_1(0) \right\rangle_S {=}
\left\langle \mathcal{O}_1(r) \mathcal{O}_1(0) \right\rangle_0 +
\delta^{(1)}_{SR} \left\langle \mathcal{O}_1(r) \mathcal{O}_1(0) \right\rangle \nn \\
&& \ \ \ \ \ \ \ \
{+}\delta^{(1)}_{LR} \left\langle \mathcal{O}_1(r) \mathcal{O}_1(0) \right\rangle.
\end{eqnarray}
The leading term in Eq.~(\ref{eq:two-point-0}) gives the two-point function of the pure system:
\begin{eqnarray} \label{eq:two-point-0A}
&& \left\langle \mathcal{O}_1(r) \mathcal{O}_1(0) \right\rangle_0 = \left\langle \sin \sqrt{\pi} \phi_1(r) \sin \sqrt{\pi} \phi_1(0) \right\rangle_0 \nn \\
&&  =  \frac{1}{(2 i)^2} \left\langle (e^{i \sqrt{\pi} \phi_1(r)} - e^{-i \sqrt{\pi} \phi_1(r)}  ) \right. \nn \\
&& \times \left. (e^{i \sqrt{\pi} \phi_1(0)} - e^{-i \sqrt{\pi} \phi_1(0)})  \right\rangle_0 \nn \\
&&  =  \frac{1}{2} \left\langle e^{i \sqrt{\pi} (\phi_1(r)-\phi_1(0))}
\right\rangle_0 = \frac12 (\Lambda r)^{-1/2},
\end{eqnarray}
where we used Eqs.~(\ref{eq:oo-cor1}) and (\ref{eq:oo-cor2}).
The first-order correction in the SR correlated disorder has been calculated in Ref.~\cite{shankar87} using bosonization
of the 2D massive Thirring model~\cite{zinn-justin-book}.
The latter allows one to eliminate the terms in action~(\ref{eq:action-sine-Gordon-2})
which are diagonal in replicas and local in space
by means of the identity
\begin{eqnarray} \label{eq:bosonosation-identity}
&&  \left[ \frac{\Lambda}{\pi} \cos \sqrt{4\pi} \varphi (r) \right]^2 = - \frac1{2\pi}  (\nabla \varphi)^2.
\end{eqnarray}
As a result, the kinetic term is rescaled by the factor of
$1+u_0/(2\pi)$. The non-diagonal in replicas terms do not contribute
to the one-loop order. Using the rescaling $\varphi=
[1+u_0/(2\pi)]^{-1/2} \varphi' $ we obtain for the SR disorder
\begin{eqnarray} \label{eq:two-point-00}
&&  \left\langle \mathcal{O}_1(r) \mathcal{O}_1(0) \right\rangle_0 + \delta^{(1)}_{SR} \left\langle \mathcal{O}_1(r) \mathcal{O}_1(0) \right\rangle
 \nn \\
&& = \frac12 (\Lambda r)^{-1/[2 (1+u_0/(2\pi))]}
\approx \frac12 (\Lambda r)^{-1/2}\left [ 1+ \frac{u_0 \ln r\Lambda }{4\pi}   \right]. \nn \\
\end{eqnarray}
For the LR disorder we calculate the correction explicitly:
\begin{eqnarray} \label{eq:two-point-000}
&& \delta^{(1)}_{LR} \left\langle \mathcal{O}_1(r) \mathcal{O}_1(0) \right\rangle = \frac{\Lambda^2}{2\pi^2} \int d^2 r_1 d^2 r_2 g(r_1-r_2)
  \nn \\
&& \left\langle \sin \sqrt{\pi} \phi_1(r)\times \sin \sqrt{\pi} \phi_1(0) \cos  \sqrt{4\pi} \varphi_1(r_1)   \right. \nn \\
&&  \left. \cos \sqrt{4\pi} \varphi_1(r_2) \right\rangle  =  - \frac{\Lambda^2}{2\pi^2} \frac{2^2}{2^2(2 i)^2} \int d^2 r_1 d^2 r_2 g(r_1-r_2) \nn \\
&& \times \left\langle e^{i \sqrt{\pi} \phi_1(r)} e^{-i \sqrt{\pi} \phi_1(0)} e^{i \sqrt{4\pi} \phi_1(r_1)} e^{-i \sqrt{4\pi} \phi_1(r_2)}  \right\rangle \nn \\
&&  =  \frac{\Lambda^2}{8\pi^2} (\Lambda r)^{-1/2} \int d^2 r_1 d^2 r_2 g(r_1-r_2) (\Lambda |r_1-r_2| )^{-2} \nn \\
&& \times \frac{|r-r_1| |r_2|}{|r-r_2| |r_1|}.
\end{eqnarray}
Taking $g(r_1-r_2)$ as the inverse Fourier transform of (\ref{furier}) at $d=2$,
\begin{eqnarray} \label{eq:dis-corr-real-sp}
g(r_1-r_2) = u_0\delta^{(2)}(r_1-r_2) + \frac{v_0 \delta}{2\pi}|r_1-r_2|^{-a},
\end{eqnarray}
where $\delta^{(2)}$  is the two-dimensional $\delta$-function, and
setting $u_0=0$, we find
\begin{eqnarray} \label{eq:two-point-2}
&& \delta^{(1)}_{LR} \left\langle \mathcal{O}_1(r) \mathcal{O}_1(0) \right\rangle =
 \frac{(\Lambda r)^{-1/2}}{8\pi^2}  \frac{v_0 \delta}{2\pi} |r|^{\delta} J\left(\frac12, \frac{\delta}{4}\right),\ \ \ \ \
\end{eqnarray}
where we have introduced the integral
\begin{eqnarray} \label{eq:two-point-3}
J(p,\tau) =
\mathcal{ FP} \int d^2 r_1 d^2 r_2 |r_1-r_2|^{4(\tau-1)} \left [\frac{|e-r_1| |r_2|}{|e-r_2| |r_1|} \right]^{2p}. \nn \\
\end{eqnarray}
Here $e$ is an arbitrary unit vector, and $\mathcal{ FP}$
means "finite part" in the sense of dimensional regularization.
The method of computing integrals of type~(\ref{eq:two-point-3}) has
been developed in Refs.~\cite{Dotsenko-int-95,Guida1998,Zoran2013}.
It reads
\begin{eqnarray} \label{eq:two-point-31}
\frac{J(p,\tau)}{4 \pi^2} = \frac{p^2}{8\tau^2} + O(\tau^{-1}).
\end{eqnarray}
Collecting all factors we arrive at Eq.~(\ref{eq:two-point-6}).


\end{document}